\documentclass[12pt,aps,nofootinbib,superscriptaddress,preprintnumbers]{revtex4-1}

\usepackage{amsmath,amssymb,bm,hyperref}
\usepackage[dvips]{graphicx}
\usepackage{yfonts}
\newcommand{\beq}{\begin{eqnarray}}
\newcommand{\eeq}{\end{eqnarray}}
\renewcommand\d{\partial}
\newcommand{\tr}{\mathop{\mathrm{tr}}}
\begin{document}

\preprint{EFI-12-29, INT-PUB-12-051, YITP-12-95}
\title{Kinetic theory with Berry curvature from quantum field theories}
\author{Dam Thanh Son}
\affiliation{Enrico Fermi Institute, University of Chicago, 5640 South
Ellis Avenue, Chicago, Illinois 60637, USA}
\author{Naoki Yamamoto}
\affiliation{Yukawa Institute for Theoretical Physics,
Kyoto University, Kyoto 606-8502, Japan}
\affiliation{Maryland Center for Fundamental Physics, 
Department of Physics, University of Maryland,
College Park, Maryland 20742-4111, USA}
\begin{abstract}
A kinetic theory can be modified to incorporate triangle anomalies and 
the chiral magnetic effect by taking into account the Berry curvature flux 
through the Fermi surface. We show how such a kinetic theory can be 
derived from underlying quantum field theories.
Using the new kinetic theory, we also compute the parity-odd correlation 
function that is found to be identical to the result in the perturbation 
theory in the next-to-leading order hard dense loop approximation.
\end{abstract}
\maketitle

\section{Introduction}
Kinetic theory \cite{Landau_kinetics} has wide applications 
in condensed matter physics, nuclear physics, astrophysics, and cosmology.
There is, however, a key deficiency in the conventional relativistic kinetic framework: 
it misses the effect of triangle anomalies \cite{Adler, BellJackiw}---an important 
feature of relativistic quantum field theories.
Recently it has been shown in Ref.~\cite{Son:2012wh}%
\footnote{See Refs.~\cite{Kirilin:2012sd, Zahed:2012yu, Son:2012bg, Gorsky:2012gi, 
Stephanov:2012ki} for further investigations and applications. 
See also Refs.~\cite{Loganayagam:2012pz, Gao:2012ix} for different 
approaches to derive kinetic equations with triangle anomalies without
referring to the Berry curvature.}
that a kinetic theory for Fermi liquids can be modified to include such anomalous 
effects by taking into account the Berry phase and Berry curvature \cite{Berry}---the 
notions extensively studied and widely applied in condensed matter 
physics~\cite{Xiao:2010}. It was shown that not only the form of the transport 
equation but also the definition of the particle number current must be modified 
when the Berry curvature has a nonzero flux through the Fermi surface.
A consequence of this modification is the generation of parity-violating and dissipationless 
current in the presence of magnetic field called the chiral magnetic effect
\cite{Vilenkin:1980fu, Nielsen:1983rb, Alekseev:1998ds, Fukushima:2008xe}.
This had been previously found in the perturbation theory \cite{Vilenkin:1980fu, Fukushima:2008xe} 
and the gauge/gravity duality \cite{Erdmenger:2008rm, Banerjee:2008th} and was
incorporated in the framework of hydrodynamics \cite{Son:2009tf} 
(see also Refs.~\cite{Banerjee:2012iz, Jensen:2012jy} for more recent developments).
The chiral magnetic effect may have been experimentally observed in relativistic 
heavy ion collisions \cite{Fukushima:2008xe, Kharzeev:2010gr} and is potentially
observable in Weyl semimetals which possess band-touching points 
\cite{Vishwanath, BurkovBalents, Xu-chern}.

On the other hand, one should be able to derive the kinetic theories from 
the underlying quantum field theories by following the standard procedure: 
starting from the equations of motion for the two-point function 
$\langle \psi(x) \psi^{\dag}(y) \rangle$ and performing a derivative expansion 
for its gauge-covariant Wigner transform, 
one arrives at the Vlasov equation (see, e.g., Ref.~\cite{Blaizot:2001nr} for a review).
So far, Berry curvature corrections to the relativistic kinetic theory 
have been ignored in the field theoretic derivation.
Also, microscopic origin of the modification to the particle number current 
is not yet clear.

In this paper, we microscopically derive the kinetic theory with Berry
curvature corrections from underlying quantum field theories.%
\footnote{See Refs.~\cite{ShindouBalents, WongTserkovnyak} for related attempts.}
For concreteness, we consider the system of relativistic chiral fermions at 
finite chemical potential $\mu$ (which is known to have a nonzero Berry curvature flux).
Our starting point is the high density effective theory \cite{Hong:1998tn, Schafer:2003jn}
that describes the physics near the Fermi surface of chiral fermions. 
In this effective theory, 
one decomposes two-component chiral fermions into single-component particles 
$\psi_+$ with positive energy $E = |{\bf p}|-\mu$ 
and antiparticles $\psi_-$ with negative energy $E = -|{\bf p}|-\mu$. 
Then one usually concentrates on the former with $E \sim 0$ for
$|{\bf p}| \sim \mu$, while neglecting the latter with $E \sim -2\mu$;
picking up only $\psi_+$ degrees of freedom leads to the conventional Vlasov equation.
As we shall demonstrate in this paper, however, if one carefully 
integrates out $\psi_-$ degrees of freedom, Berry curvature corrections 
emerge in the kinetic theory from the mixing between 
$\psi_+$ and $\psi_-$ (or $\psi_-$ and $\psi_-$). The modification to 
Liouville's theorem on the phase space known in the condensed matter 
literature \cite{Xiao:2005, Duval:2005} and the modification
to the current found in Ref.~\cite{Son:2012wh}
can be naturally understood from this deliberate integrating out procedure
[see Eqs.~(\ref{eq:n_micro}) and (\ref{eq:j_micro})].
Apparently, the essential ingredient in this field theoretic argument to 
lead to Berry curvature corrections is a Fermi surface of chiral fermions.

We also compute the parity-violating correlation function using the
kinetic theory with Berry curvature corrections. In the case of the conventional 
Vlasov equation, it is known that the parity-even correlation function
computed in the kinetic theory coincides with the one in the perturbation 
theory under the hard dense loop approximation \cite{Blaizot:2001nr, Manuel:1995td}. 
In this paper, we will see that the parity-odd correlation function 
derived from the new kinetic theory is equivalent to the result in the 
perturbation theory beyond the leading-order hard dense loop approximation. 

The paper is organized as follows. In Sec.~\ref{sec:Berry}, we review 
the kinetic theory with Berry curvature corrections. We also derive a 
new relation for the spin magnetic moment of quasiparticles in 
Fermi liquids. In Sec.~\ref{sec:qft}, we derive the new kinetic theory 
starting from quantum field theories.
In Sec.~\ref{sec:correlation}, we compute the parity-violating 
correlation functions using both the new kinetic theory and 
perturbation theory and confirm their agreement.
Section~\ref{sec:conclusion} is devoted to our conclusions.

Throughout the paper, we consider sufficiently low temperature
regime $T \ll \mu$ where the Fermi surface is well defined.
We also concentrate on the collisionless limit of the kinetic theory.

\section{Kinetic theory with Berry curvature}
\label{sec:Berry}
In this section, we review the kinetic theory in the presence of 
the Berry curvature which exhibits triangle anomalies \cite{Son:2012wh} 
(see also Refs.~\cite{Son:2012bg, Stephanov:2012ki})
and provide the proper definitions of particle number density and current. 
We also derive the dispersion relation of quasiparticles according to
the constraints of Lorentz invariance.

\subsection{Berry curvature and Poisson brackets}
We first consider a single chiral fermion expressed by 
the two-component spinor $u_{\bf p}$ satisfying the Weyl equation
\beq
({\bm \sigma} \cdot {\bf p}) u_{\bf p} = \pm |{\bf p}| u_{\bf p},
\eeq
where the signs $+$ and $-$ correspond to 
right-handed and left-handed fermions, respectively.
The two-component spinor described above has a 
nonzero Berry connection defined by \cite{Berry}
\beq
\label{eq:phase}
i \bm{\mathcal A}_{\bf p}  \equiv 
u_{\bf p}^{\dag} \bm{\nabla}_{\bf p} u_{\bf p},
\eeq
and a nonzero Berry curvature,
\beq
\label{eq:curvature}
\bm {\Omega}_{\bf p} \equiv \bm{\nabla}_{\bf p} \times \bm{\mathcal A}_{\bf p}
= \pm \frac{\hat {\bf p}}{2|{\bf p}|^2},
\eeq
where $\hat {\bf p}={\bf p}/|{\bf p}|$ is a unit vector.
Equations (\ref{eq:phase}) and (\ref{eq:curvature}) can be regarded 
as the fictitious vector potential and magnetic field
in the momentum space.
This fictitious magnetic field can be associated with the one
from a ``magnetic monopole" with the charge $\pm 1/2$ 
put in the center of the momentum space. 
As a result, the motion of chiral fermions is affected by 
the Berry curvature in the momentum space, in addition to 
the usual electromagnetic fields in the coordinate space.
In particular, the effects of the Berry curvature work 
oppositely between right-handed and left-handed chiral fermions.

Let us now consider the action of a single quasiparticle in the presence
of the electromagnetic fields and Berry curvature \cite{Xiao:2005, Duval:2005}, 
\beq
\label{eq:S}
S = \int dt[p^i \dot x^i  + A^i(x) \dot x^i - 
{\cal A}^i(p) \dot p^i - \epsilon_{\bf p}(x) -  A^0(x)].
\eeq
Note that the quasiparticle energy $\epsilon_{\bf p}$ is a function of $x$ in general; 
indeed chiral fermions have the magnetic moment at finite chemical potential $\mu$ 
and their energy depends on the magnetic field ${\bf B}(x)$ 
[see Eq.~(\ref{eq:dispersion_Lorentz}) below].
The action (\ref{eq:S}) can be summarized in the following form, 
by combining space $x$ and momentum $p$ into a set of 
variables $\xi^a$ ($a=1$, $\cdots$, 6),
\beq
\label{eq:action}
S = \int dt[-\omega_a(\xi)\dot \xi^a - H(\xi)],
\eeq
where $H(\xi)=\epsilon_{\bf p} + A_0$ is the Hamiltonian.

The equations of motion of the action (\ref{eq:action}) read
\beq
\omega_{ab}\dot \xi^b= -\d_a H,
\eeq
where $\omega_{ab}=\d_a \omega_b - \d_b \omega_a$ and $\d_a \equiv \d/\d \xi^a$.
This equation can be rewritten as
\beq
\label{eq:eom}
\dot \xi^a = - \omega^{ab}\d_b H,
\eeq
where $\omega^{ab} \equiv (\omega^{-1})^{ab}$ is the inverse
matrix of $\omega_{ab}$. Here we assume the existence of the
inverse matrix, i.e., $\omega \equiv \det\omega_{ab} \neq 0$.
Equation (\ref{eq:eom}) can be interpreted as 
\beq
\dot\xi^a = \{H,\, \xi^a\} = - \{\xi^a,\, \xi^b\} \frac{\d H}{\d \xi^b}\,,
\eeq
once we define the Poisson brackets as
\beq
\{ \xi_a, \xi_b \} = \omega^{ab}.
\eeq
The explicit forms of the Poisson brackets for the action (\ref{eq:action}) read
\cite{Duval:2005}
\beq
\label{PB}
  \{ p_i,\, p_j\} = - \frac{\epsilon_{ijk} B_k}{1+{\bf B}\cdot\bm{\Omega}}\,, \quad
  \{ x_i,\, x_j\} = \frac{\epsilon_{ijk}\Omega_k}{1+{\bf B}\cdot\bm{\Omega}}
   \,, \quad
  \{ p_i,\, x_j\} = \frac{\delta_{ij} + \Omega_i B_j}
  {1+{\bf B}\cdot\bm{\Omega}}\,,
\eeq
where $B^i=\epsilon^{ijk}\d A^k/\d x^j$.

These Poisson brackets should be compared with the usual ones
in the absence of the Berry curvature and electromagnetic fields
\beq
  \{ p_i,\, p_j\} = 0 \,, \quad
  \{ x_i,\, x_j\} = 0 \,, \quad
  \{ p_i,\, x_j\} = \delta_{ij} \,,
\eeq
whose invariant phase space is $d{\bf p}d{\bf x}/(2\pi)^3$.
As a consequence of the modifications to the Poisson brackets above, 
the invariant phase space is modified to \cite{Xiao:2005}
\beq
\label{eq:phase-space}
 {\rm d}\Gamma = \sqrt{\omega}\, {\rm d}\xi
  =(1+ {\bf B}\cdot\bm{\Omega}_{\bf p}) 
  \frac{{\rm d}{\bf p}\,{\rm d}{\bf x}}{(2\pi)^3} \,,
\eeq
where $\omega \equiv \det\omega_{ab}$.

\subsection{Kinetic theory with Berry curvature and triangle anomalies}
\label{sec:kinetic}
Let us construct the collisionless kinetic theory incorporating
the effects of the Berry curvature.
If collisions between particles are negligible, 
each particle constitutes a closed subsystem.
According to Liouville's theorem, which states that a volume element in the 
phase space does not change during its time evolution, the one-particle 
distribution function $n(\xi)$ would obey $d n/dt=0$. However, the invariant 
phase space is modified as Eq.~(\ref{eq:phase-space}) due to the Berry curvature, 
and the probability of finding a particle in the phase space is 
$\sqrt{\omega}n(\xi) d {\xi}$.
As a result, we instead use the modified distribution function 
$\rho(\xi) = \sqrt{\omega} n(\xi)$ that obeys the equation $d \rho/dt=0$,
or equivalently,
\beq
\dot \rho + \d_a (\dot \xi^a \rho)=0.
\eeq 
Using Eq.~(\ref{eq:eom}), this reduces to
\beq
\label{eq:kinetic_sum}
\dot n_{\bf p} - \omega^{ab}\d_b H \d_a n_{\bf p} = 0.
\eeq
Setting $H = \epsilon_{\bf p} + A_0$, we can explicitly write down 
the kinetic equation 
(see also Refs.~\cite{Duval:2005, Son:2012bg, Stephanov:2012ki})
\beq
\label{eq:kinetic}
\dot n_{\bf p} + \frac{1}{1 + {\bf B} \cdot {\bf \Omega}_{\bf p}}
\left[\left(\tilde {\bf E} + \tilde {\bf v} \times {\bf B}
+ (\tilde {\bf E} \cdot {\bf B}) {\bm \Omega}_{\bf p} \right)
\cdot \frac{\d n_{\bf p}}{\d {\bf p}}
+ \left(\tilde {\bf v} + \tilde {\bf E} \times {\bm \Omega}_{\bf p}
+ (\tilde {\bf v} \cdot {\bm \Omega}_{\bf p}) {\bf B}
\right) \cdot \frac{\d n_{\bf p}}{\d {\bf x}}
\right]=0,
\nonumber \\
\eeq
where $\tilde {\bf v} = \d{\epsilon_{\bf p}}/\d {\bf p}$ and
${\tilde {\bf E}}={\bf E} - \d{\epsilon_{\bf p}}/\d {\bf x}$.
This is a general low-energy effective theory in the presence of
Berry curvature corrections that describes the evolution of $n_{\bf p}$.
Note that $\tilde {\bf v}$ is different from the unit vector $\hat {\bf p}$
when the quasiparticle energy $\epsilon_{\bf p}$ has the contribution 
from the magnetic moment [see Eq.~(\ref{eq:dispersion_Lorentz}) below].
If we turn off the Berry curvature 
and ignore the ${\bf x}$ dependence of $\epsilon_{\bf p}$
(i.e., ${\bf \Omega}_{\bf p}=0$ and $\d{\epsilon_{\bf p}}/\d {\bf x}=0$), 
this reduces to the usual Vlasov equation. 

We now define the particle number density
\beq
\label{eq:n}
n = \int \frac{d^3p}{(2\pi)^3}(1+{\bf B}\cdot{\bm \Omega}_{\bf p})n_{\bf p},
\eeq
where the invariant phase space is modified according to Eq.~(\ref{eq:phase-space}).
Multiplying the kinetic equation (\ref{eq:kinetic}) by $\sqrt{\omega}$, performing 
the integral over momentum ${\bf p}$, and using Maxwell equations, 
${\bm \nabla} \cdot {\bf B}=0$ and $\d_t {\bf B} + {\bm \nabla} \times {\bf E}=0$,
we obtain the following identity \cite{Son:2012wh}:
\beq
\label{eq:anomaly0}
\d_t n + {\bm \nabla} \cdot {\bf j}
= -\int \frac{d^3p}{(2\pi)^3} \left({\bm \Omega}_{\bf p} \cdot \frac{\d n_{\bf p}}{\d {\bf p}} \right) 
{\bf E} \cdot {\bf B},
\eeq
where
\beq
\label{eq:j}
{\bf j} = -\int \frac{d^3p}{(2\pi)^3}
\left[\epsilon_{\bf p}\frac{\d n_{\bf p}}{\d {\bf p}}
+ \left({\bm \Omega}_{\bf p} \cdot \frac{\d n_{\bf p}}{\d {\bf p}}\right) \epsilon_{\bf p} {\bf B}
+ \epsilon_{\bf p} {\bm \Omega}_{\bf p} \times \frac{\d n_{\bf p}}{\d {\bf x}} \right] 
+ {\bf E} \times {\bm \sigma}, 
\eeq
is identified with the current and ${\bm \sigma}$ is defined as
\beq
\label{eq:sigma}
{\bm \sigma} = \int \frac{d^3p}{(2\pi)^3} {\bm \Omega}_{\bf p} n_{\bf p}.
\eeq
In Eq.~(\ref{eq:anomaly0}), we observe that
the particle number of chiral fermions is no longer conserved
when we turn on both electric and magnetic fields.
By integration by part and using ${\bm \nabla}_{\! \! \bf p} \cdot {\bm \Omega}_{\bf p}=0$ 
around the Fermi surface, $n_{\bf p}=1$ deep inside the Fermi surface 
and $n_{\bf p}=0$ far outside the Fermi surface, it can be evaluated as
\beq
\label{eq:anomaly}
\d_t n + {\bm \nabla} \cdot {\bf j} = \pm \frac{1}{4\pi^2}{\bf E} \cdot {\bf B},
\eeq
for right-handed and left-handed fermions, respectively.
This is exactly the equation of triangle anomalies in relativistic quantum field theories 
\cite{Adler, BellJackiw}, which holds independently of interactions.

The first term in Eq.~(\ref{eq:j}) is the usual particle number current
of the kinetic theory, while the remaining terms are the Berry curvature
corrections. The same form of the current can be obtained in the Hamiltonian 
formalism using the commutation relations postulated in Ref.~\cite{Son:2012wh}.
The final term in Eq.~(\ref{eq:j}) is the anomalous Hall current,
which vanishes for a spherically symmetric distribution function at rest. 
In this case, the current is
\beq
\label{eq:j_sp}
{\bf j} = -\int \frac{d^3p}{(2\pi)^3}
\left[\epsilon_{\bf p}\frac{\d n_{\bf p}}{\d {\bf p}}
+ \left({\bm \Omega}_{\bf p} \cdot \frac{\d n_{\bf p}}{\d {\bf p}}\right) \epsilon_{\bf p} {\bf B}
+ \epsilon_{\bf p} {\bm \Omega}_{\bf p} \times \frac{\d n_{\bf p}}{\d {\bf x}} \right].
\eeq
At this moment, microscopic origins of the Berry curvature corrections 
to the particle number density and current in Eqs.~(\ref{eq:n}) and (\ref{eq:j_sp}) 
are not so clear. In Sec.~\ref{sec:qft}, microscopic meanings of these corrections will 
be clarified in the field theoretic language.

It should be remarked that there is an ambiguity to define the number current 
from the continuity equation because 
$\tilde {\bf j} = {\bf j} + {\bm \nabla} \times {\bf a}$ with any vector ${{\bf a}}$
is also a solution to the continuity equation. 
In order to fix this ambiguity, we look at the  energy and momentum conservations. 
We define the energy density and the momentum density, 
\beq
\epsilon = \int \frac{d^3p}{(2\pi)^3} (1+{\bf B} \cdot {\bf \Omega}_{\bf p})
 \epsilon_{\bf p} n_{\bf p}, 
\qquad
\pi^i = \int \frac{d^3p}{(2\pi)^3} (1+{\bf B} \cdot {\bf \Omega}_{\bf p})
p^i n_{\bf p}.
\label{eq:momentum_density}
\eeq
Multiplying Eq.~(\ref{eq:kinetic}) by $\epsilon_{\bf p} \sqrt{\omega}$ 
and $p^i \sqrt{\omega}$, and performing the integral over momentum ${\bf p}$, 
we have
\beq
\d_t \epsilon + \int \frac{d^3p}{(2\pi)^3} \epsilon_{\bf p} F=0, \qquad
\d_t \pi^i + \int \frac{d^3p}{(2\pi)^3} p^i F=0,
\eeq
where $F$ is the piece in the square brackets of Eq.~(\ref{eq:kinetic}).
The above equations can be respectively interpreted as the energy and 
momentum conservation laws
\beq
\d_{\mu} T^{0 \mu} = E^i j^i,  \qquad
\d_{\mu} T^{i \mu} = n E^i + \epsilon^{ijk}j^j B^k,
\eeq
where
\begin{align}
\label{eq:energy_flux}
T^{0i} &= - \int \frac{d^3p}{(2\pi)^3} \left[(\delta^{ij} + B^i \Omega^j)
\frac{\epsilon_{\bf p}^2}{2} \frac{\d n_{\bf p}}{\d p^j} 
+ \epsilon^{ijk}\frac{\epsilon_{\bf p}^2}{2}
\Omega^j \frac{\d n_{\bf p}}{\d x^k} \right], 
\\
T^{ij} &= -\int \frac{d^3p}{(2\pi)^3} p^i 
\left[\epsilon_{\bf p}(\delta^{jk} + B^j\Omega^k) \frac{\d n_{\bf p}}{\d p^k} 
+ \epsilon^{jkl} \Omega^k \left(E^l n_{\bf p} +
\epsilon_{\bf p} \frac{\d n_{\bf p}}{\d x^l}\right) \right] - \delta^{ij}\epsilon,
\end{align}
which indicates that $j^i$ is the genuine current.

Alternatively, this ambiguity is avoided if we use the definition 
of the current \cite{Son:2012bg, Stephanov:2012ki}
\beq
{\bf j} = \int \frac{d^3p}{(2\pi)^3} \sqrt{\omega} \dot {\bf x}.
\eeq
By using Eq.~(\ref{eq:eom}), one can actually check that 
this current is equal to Eq.~(\ref{eq:j}).

In equilibrium where $n_{\bf p}$ is homogeneous, the first and third terms 
in the right-hand side of Eq.~(\ref{eq:j_sp}) vanish, while 
the second term is nonvanishing. Using $\epsilon_{\bf p} = \mu$ 
at the Fermi surface with $\mu$ the chemical potential, we find
\beq
\label{eq:j_CME}
{\bf j} = - \int \frac{d^3p}{(2\pi)^3} 
\left({\bm \Omega}_{\bf p} \cdot \frac{\d n_{\bf p}}{\d {\bf p}}\right) \mu {\bf B}
= \pm \frac{\mu}{4\pi^2} {\bf B}.
\eeq
This is the relation of the chiral magnetic effect 
\cite{Vilenkin:1980fu, Nielsen:1983rb, Fukushima:2008xe}: the equilibrium current 
induced in the direction of the magnetic field for chiral fermions 
at finite chemical potential $\mu$.

\subsection{Lorentz invariance in Fermi liquids}
\label{sec:magnetic-moment}
Here we consider the consequences of Lorentz invariance in a system
described by Landau's Fermi liquid theory \cite{Landau}. The constraint due to
Lorentz invariance is that the energy flux is equal to the momentum density,
$T^{0i}=\pi^i$. From Eqs.~(\ref{eq:momentum_density}) and (\ref{eq:energy_flux}), 
this condition in the homogeneous system becomes
\beq
\label{eq:Lorentz}
- \int \frac{d^3p}{(2\pi)^3} (\delta^{ij} + B^i \Omega^j_{\bf p}) 
\frac{\epsilon_{\bf p}^2}{2} \frac{\d n_{\bf p}}{\d p^j} 
= \int \frac{d^3p}{(2\pi)^3} \ (1+{\bf B} \cdot {\bf \Omega}_{\bf p})p^i n_{\bf p}.
\eeq
We vary both sides of Eq.~(\ref{eq:Lorentz}) as
$n_{\bf p}=n_{\bf p}^0 + \delta n_{\bf p}$ and
$\epsilon_{\bf p} = \epsilon_{\bf p}^0 + \delta \epsilon_{\bf p}$, where
\beq
\label{eq:Landau}
\delta \epsilon_{\bf p} = \int \frac{d^3q}{(2\pi)^3}\ (1+{\bf B} \cdot {\bf \Omega}_{\bf q}) 
f({\bf p, q}) \delta n_{\bf q},
\eeq
with $f({\bf p, q})$ being some function characterizing the interactions 
among quasiparticles, called the Landau parameters. 
By integration by parts, we have
\beq
\label{eq:variation}
\int \frac{d^3p}{(2\pi)^3} \ (\delta^{ij} + B^i \Omega^j_{\bf p})
\left[\frac{1}{2}\frac{\d (\epsilon^0_{\bf p})^2}{\d p^j} \delta n_{\bf p}
- \epsilon^0_{\bf p} \delta \epsilon_{\bf p} \frac{\d n_{\bf p}^0}{\d p^j} \right]
= \int \frac{d^3p}{(2\pi)^3} \ (1+{\bf B} \cdot {\bf \Omega}_{\bf p})p^i \delta n_{\bf p}.
\eeq
Using Eq.~(\ref{eq:Landau}) and renaming variables ${\bf q} \leftrightarrow {\bf p}$,
the second part of the left-hand side reduces to
\beq
- \int \frac{d^3p}{(2\pi)^3} \ (1+{\bf B} \cdot {\bf \Omega}_{\bf p})
\left( \int \frac{d^3q}{(2\pi)^3} \ (\delta^{ij} + B^i \Omega^j_{\bf q})
f({\bf p,q}) \epsilon^0_{\bf q} \frac{\d n_{\bf q}^0}{\d q^j} \right) \delta n_{\bf p}.
\eeq
Because $\delta n_{\bf p}$ is arbitrary, we have the relation
\beq
\label{eq:relation}
(\delta^{ij} + B^i \Omega^j_{\bf p}) \epsilon^0_{\bf p}
\frac{\d \epsilon^0_{\bf p}}{\d p^j}
- (1+{\bf B} \cdot {\bf \Omega}_{\bf p})
\int \frac{d^3q}{(2\pi)^3} (\delta^{ij} + B^i \Omega^j_{\bf q})
f({\bf p,q}) \epsilon^0_{\bf q} \frac{\d n_{\bf q}^0}{\d q^j}
= (1+{\bf B} \cdot {\bf \Omega}_{\bf p})p^i. \nonumber \\
\eeq

To proceed, we take an ansatz 
\beq
\label{eq:ansatz_e}
\epsilon_{\bf p}^0 = v_f (p - p_f) + \mu + \gamma(p) {{\bf B} \cdot {{\bf p}}}
\eeq
to the linear order in ${\bf B}$,
with $v_f$ and $p_f$ being some constants and 
$\gamma$ a scalar function of $p \equiv |{\bf p}|$.
Note that the Fermi velocity defined by $\d \epsilon_{\bf p}^0/\d p$ is 
${\bf B}$ dependent. Note also that Landau parameters are functions of ${\bf B}$; 
from the property $f({\bf p, q}) = f({\bf q, p})$, Landau parameters are composed of
two parts (to the linear order in ${\bf B}$),
\beq
\label{eq:ansatz_f}
f({\bf p, q}) = f^A({\bf p, q}) +  
\frac{{\bf B} \cdot (\hat {\bf p}+{\hat {\bf q})}}{2p_f^2} f^B({\bf p, q}),
\eeq
where $f^{A,B}({\bf p, q})$ are independent of ${\bf B}$ and can be expanded 
by the Legendre functions as
\beq
f^{A,B}({\bf p, q}) = \sum_{l=0}^{\infty} f_l^{A,B} P_l (\cos \theta),
\eeq
where $\theta$ is the angle between ${\bf p}$ and ${\bf q}$ 
both taken on the Fermi surface.

We now evaluate both sides of Eq.~(\ref{eq:relation}) to the
linear order in ${\bf B}$. Substituting Eqs.~(\ref{eq:ansatz_e}) and (\ref{eq:ansatz_f}),
replacing $p$ by $p_f$, and performing the angular integral
(note also that $n_{\bf p}^0$ has the ${\bf B}$ dependence), we have
\beq
\label{eq:relation2}
& & \mu v_f \left(1 + \frac{1}{3}F_1^A \right) \hat p^i
+ \left[p_f \left(v_f \gamma + \mu \gamma' - \frac{1}{3} \mu v_f F_1'^A \right)
+ \frac{\mu v_f}{2p_f^2} \left(\frac{1}{3}F_1^A + \frac{1}{3}F_1^B + \frac{1}{5}F_2^B \right) \right]
({\bf B} \cdot {\hat {\bf p}}) \hat p^i \nonumber \\
& & \ 
+ \mu \left[\frac{v_f}{2p_f^2}\left(1 + F_0^A + \frac{1}{3}F_0^B -\frac{1}{15}F_2^B \right) + \gamma \right] B^i 
= \left(p_f + \frac{{\bf B} \cdot {\bf {\hat p}}}{2p_f} \right) {\hat p}^i,
\eeq
where $\gamma \equiv \gamma(p_f)$, $\gamma'\equiv \frac{\d}{\d p}\gamma(p_f)$, and we defined 
\beq
\int d {\hat {\bf q}} \frac{\d f^A({\bf p,q})}{\d q^i}({\bf B} \cdot {\bf {\hat q}})
\equiv  \frac{1}{3} f_1'^A ({\bf B} \cdot {\bf {\hat p}}) \hat p^i .
\eeq
In order to satisfy Eq.~(\ref{eq:relation2}) for any ${\bf B}$ and ${\hat {\bf p}}$, 
we must have
\begin{gather}
\label{eq:Baym-Chin}
v_f \left(1 + \frac{1}{3}F_1^A \right) = \frac{p_f}{\mu}, \\
\label{eq:mag_moment}
\gamma = -\frac{v_f}{2p_f^2}
\left(1 + F_0^A + \frac{1}{3}F_0^B - \frac{1}{15}F_2^B \right), \\
\label{eq:derivative}
p_f \left(v_f \gamma + \mu \gamma' - \frac{1}{3} \mu v_f F_1'^A \right)
+ \frac{\mu v_f}{2p_f^2} \left(\frac{1}{3}F_1^A + \frac{1}{3}F_1^B + \frac{1}{5}F_2^B\right) = \frac{1}{2p_f}
\end{gather}
While Eq.~(\ref{eq:Baym-Chin}) is the relation obtained by 
Baym and Chin \cite{Baym:1975va}, Eq.~(\ref{eq:mag_moment}) 
is a new relation for the anomalous spin magnetic moment. 
A constraint from the gauge invariance on the anomalous 
{\it angular} magnetic moment in Fermi liquids was originally
given by Migdal \cite{Migdal} and was studied in detail in Ref.~\cite{Bentz:1985qh}.
Here we have provided the new constraint on the anomalous 
{\it spin} magnetic moment from the viewpoint of the Berry curvature 
together with the Lorentz invariance.

In particular, in the noninteracting limit where $f({\bf p, q})$ is turned off
and $p_f=\mu$, we obtain
\beq
v_f=1, \qquad \gamma(\mu) = -\frac{1}{2\mu^2}, \qquad \gamma'(\mu)=\frac{1}{\mu^3}.
\eeq
A solution to satisfy these relations is taken as 
$\gamma(p) = -1/(2p^2)$. In this case, we have
\beq
\label{eq:dispersion_Lorentz}
\epsilon_{\bf p}^0 = p - \frac{{\bf B} \cdot {\bf {\hat p}}}{2p}.
\eeq 
We shall see in Sec.~\ref{sec:qft} 
[Eqs.~(\ref{eq:dispersion}) and (\ref{eq:dispersion_derivative})]
based on the microscopic quantum field theories, that this is actually 
the dispersion relation of chiral fermions near the Fermi surface
in a magnetic field.

\section{From quantum field theories to kinetic theory with Berry curvature}
\label{sec:qft}
In this section, we derive the kinetic theory constructed in Sec.~\ref{sec:kinetic} 
from the microscopic quantum field theories: the kinetic equation (\ref{eq:kinetic}) 
and the modified number density (\ref{eq:n}) and current (\ref{eq:j}) are reproduced 
microscopically. The resultant kinetic theory exhibits triangle anomalies and 
the chiral magnetic effect.

Our procedure is as follows: we first consider the high density effective theory
\cite{Hong:1998tn, Schafer:2003jn}, which is an effective field theory valid
near the Fermi surface. The expansion parameter of the theory is $l/\mu$ where
$l$ is the residual momentum measured from the Fermi surface. 
We then derive the kinetic theory by performing the derivative expansion
for the equations of motion of the Wigner function defined 
in the high density effective theory.
The expansion parameter here is the slowly varying disturbances $\d_X$
taken to be much smaller than the chemical potential $\mu$ 
and the gauge field $A_{\mu}$.
Note that our procedure does not rely on the
expansion in terms of the coupling constant, and hence, it is applicable 
even when the interactions are strong as long as the notion of quasiparticles
is well defined.

In this and following sections, we consider the theory 
with right-handed chiral fermions.

\subsection{High density effective theory}
\label{sec:HDET}
We first review the derivation of the high density effective theory 
\cite{Hong:1998tn, Schafer:2003jn}. We start with the Lagrangian
for right-handed fermions,
\beq
\label{eq:Lagrangian}
{\cal L}=\psi^{\dag} (i {\not \! \! D} + \mu)\psi,
\eeq
where ${\not \! \! D}=\sigma^{\mu} D_{\mu}$
with $D_{\mu}=\partial_{\mu} + i A_{\mu}$ and $\sigma^{\mu}=(1,{\bm \sigma})$,
and $A_{\mu}$ is the external background field. 

In order to focus on the particles near the Fermi surface, we decompose
the energy and momentum of a particle (or a hole) near the Fermi surface 
as $p^0 = \mu + l^0$ and ${\bf p}=\mu {\bf v} + {\bf l}$ with 
$l^0, |{\bf l}| \ll \mu$, where ${\bf v}$ is a unit vector which specifies 
a direction to a point on the Fermi surface. 
The momentum can be shifted by $\mu {\bf v}$ by performing a Fourier transformation
\beq
\psi(x) = \sum_{v} e^{i \mu {\bf v} \cdot {\bf x}} \psi_v(x),
\eeq
where summation is taken over ${\bf v}$. 
The matrix ${\bm \sigma} \cdot {\bf v}$ in the momentum space 
can be diagonalized by using the projectors as
\beq
\psi_{\pm v}= {P}_{\pm}({\bf v}) \psi_v, \qquad
{P}_{\pm}({\bf v})=\frac{1 \pm {\bm \sigma} \cdot {\bf v}}{2}.
\eeq
In the absence of the external electromagnetic field, $\psi_{\pm}$ 
satisfy the eigenvalue equations, 
$({\bm \sigma} \cdot {\bf v}) \psi_{\pm} = \pm \psi_{\pm}$.

In terms of $\psi_{\pm v}$, Eq.~(\ref{eq:Lagrangian}) reduces to
\beq
\label{eq:decompose}
\psi^{\dag} (i {\not \! \! D} + \mu )\psi
&=& \sum_{v}  [\psi_{+v}^{\dag}iv \cdot D \psi_{+v} + 
\psi_{-v}^{\dag}( 2\mu + i {\bar v} \cdot D) \psi_{-v}
+ (\psi_{+v}^{\dag} i {\not \! \! D}_{\perp} \psi_{-v} + {\rm h.c.})],
\eeq
where 
$v^{\mu}=(1, {\bf v})$, $\bar v^{\mu}=(1, -{\bf v})$, 
$\sigma_{\perp}^{\mu} = (0, {\bm \sigma} - {\bf v}({\bf v} \cdot {\bm \sigma}))$,
$D_{\perp}^{\mu} = (0, {\bf D} - {\bf v}({\bf v} \cdot {\bf D}))$,
and ${\not \! \! D}_{\perp} = \sigma_{\perp}^{\mu} D_{\mu} = \sigma^{\mu} D_{\mu}^{\perp}$.
By integrating out $\psi_{-v}$ using the equation of motion for $\psi_{-v}$,
\beq
(2\mu + i {\bar v} \cdot D)\psi_{-v} + i {\not \! \! D_{\perp}} \psi_{+v}=0,
\label{eq:relation+-}
\eeq
the effective Lagrangian in terms of $\psi_+$ can be written down order by order in $1/\mu$,
\begin{gather}
\label{eq:HDET}
{\cal L}_{\rm EFT} = \sum_{n} {\cal L}^{(n)}, \qquad
{\cal L}^{(n)} = \sum_{v} \psi_{+v}^{\dag} {\cal D}^{(n)} \psi_{+v}, 
\end{gather}
where ${\cal L}^{(n)}$ denotes the effective Lagrangian of the $n$ th order 
in $1/\mu$ ($n=0,1,2,\cdots$).
The explicit expressions for ${\cal D}^{(n)}$ ($n=0,1,2$) are
\beq
{\cal D}^{(0)} = iv \cdot D,
\qquad
{\cal D}^{(1)} = \frac{{\not \! \! D}_{\perp}^2}{2\mu},
\qquad
\label{eq:D^2}
{\cal D}^{(2)} = -\frac{i}{4\mu^2}
{\not \! \! D}_{\perp} (\bar v \cdot D) {\not \! \! D}_{\perp}.
\eeq

Using
${\not \! \! D}_{\perp}^2 = D_{\perp}^2 + {\bf B} \cdot {\bm \sigma}$,
$\psi^{\dag}_{+v}{\bm \sigma} \psi_{+v} = \psi^{\dag}_{+v}{\bf v} \psi_{+v}$,
and $p = \mu + l_{\parallel} + l_{\perp}^2/(2\mu) + O({1}/{\mu^2})$,
the dispersion relation near the Fermi surface reads
\beq
\label{eq:dispersion}
\epsilon_{\bf p} = p - \frac{{\bf B} \cdot {\bf v}}{2\mu} + 
O\left(\frac{1}{\mu^2} \right).
\eeq
This indeed agrees with Eq.~(\ref{eq:dispersion_Lorentz})
up to $O(1/\mu^2)$. [The agreement will be shown to the 
order of $O(1/\mu^2)$ in the next subsection.]
The second term in Eq.~(\ref{eq:dispersion}) originates from 
the magnetic moment of chiral fermions at finite $\mu$.
This is similar in structure to the Pauli equation
that describes the magnetic moment of massive Dirac fermions 
in the vacuum; the Pauli equation can be obtained by expanding
the massive Dirac equation in $1/m$, where $m$ is the
mass of Dirac fermions.

\subsection{Kinetic theory via derivative expansion}
\label{sec:blaizot-iancu}
We construct the kinetic theory based on the effective theory (\ref{eq:HDET})
on a patch indicated by a unit vector ${\bf v}$.
We consider the Dirac operator to the second order in $1/\mu$,
${\cal D}= {\cal D}^{(0)} + {\cal D}^{(1)} + {\cal D}^{(2)}$,
and introduce a two-point function for $\psi_{v}$,
\beq
G_{v}(x,y)=\langle \psi_{v}(x) \psi^{\dag}_{v}(y) \rangle.
\eeq
The function $G_{v}(x,y)$ satisfies equations of motion together with
projection conditions,
\begin{gather}
\label{eq:KB}
{\cal D}_x G_{v}(x,y)=0, \qquad G_{v}(x,y){\cal D}^{\dag}_y=0, \\
\label{eq:projection}
{P}_-({\bf v}) G_{v}(x,y)=0, \qquad G_{v}(x,y){P}_-({\bf v})=0.
\end{gather}

In thermal equilibrium where the system is homogeneous, $G_v$
depends only on the relative coordinate $s = x - y$.
We are interested in the small deviation from the equilibrium where
$G_v$ depends both on $x$ and $y$. It is thus
useful to change the coordinates from $(x,y)$ to the 
center-of-mass and relative coordinates $(X,s)$ defined by
\beq
x = X + \frac{s}{2}, \qquad y = X - \frac{s}{2},
\eeq
and consider the derivative expansion with respect to $X$.

In order to derive a quantum analogue of the classical 
distribution function, we perform the Wigner transformation
\beq
G_{v}(X,l)=\int d^4s \ e^{i l \cdot s} G_{v} 
\left(X + \frac{s}{2}, X-\frac{s}{2} \right),
\eeq
where $l^{\mu}$ is the residual four-momentum.
Unlike $G_{v}(x,y)$, however, $G_{v}(X,l)$ is not gauge covariant.
We will thus use the gauge-covariant definition instead,
\beq
\label{eq:covariant}
\tilde G_{v}(X,l) = \int d^4s \ e^{i l \cdot s} U \left(X, X + \frac{s}{2} \right)
G_{v} \left(X + \frac{s}{2}, X - \frac{s}{2} \right) U \left(X - \frac{s}{2}, X \right),
\eeq
where
\beq
U(x,y)=P \exp \left[-i \int_{\gamma} dx^{\mu} A_{\mu}(x) \right],
\eeq
is the Wilson line. The symbol $P$ is the path ordering 
along the path $\gamma$ from $x$ to $y$. For simplicity,
$\tilde G_{v}(X,l)$ is renamed $G_{v}(X,l)$ in what follows.

In constructing the kinetic theory, 
we consider the slowly varying disturbances
and perform a gradient expansion in terms of $\partial_X$. 
To this end, we assume the following counting scheme:
$\partial_X = O(\epsilon_1)$, $\partial_s = O(\epsilon_2)$, 
$A_{\mu} = O(\epsilon_3)$, and $F_{\mu \nu} =O(\epsilon_1 \epsilon_3)$.
Here $\d_X$ and $\d_s$ are the derivatives with respect to $X$ and $s$, 
and $\epsilon_i$ ($i=1,2,3$) are independent expansion parameters 
which satisfy the conditions, $\epsilon_1 \ll \epsilon_{2,3} \ll 1$. 
The condition $\epsilon_i \ll 1$ ($i=1,2,3$) is necessary 
for the derivative expansion in the high density effective theory while 
$\epsilon_1 \ll \epsilon_{2,3}$ is necessary for the derivative expansion 
in the kinetic theory.
In order to take into account triangle anomalies, 
we consider the kinetic theory to $O(\epsilon_1^2 \epsilon_3^2$).

For simplicity, in this subsection we consider the homogeneous system where 
$\d^X_{\rho} F_{\mu \nu}=0$ and $\d^X_{i} n_{\bf p}=0$ 
($n_{\bf p}$ is the distribution function which will be defined below).
This is sufficient for our purpose to understand the microscopic 
origin of the Berry curvature corrections. The generalization to the 
inhomogeneous case should be straightforward.

Consider the equations, 
${\cal D}_x G_{v}(x,y) \pm G_{v}(x,y){\cal D}_y^{\dag}=0$. We expand them
in terms of $\d_X$ to the second order and perform the Wigner transformation.
The Wigner transform of the equations can be written down order by order,
\beq
I_{\pm}^{(n)} \equiv
\int \frac{d^4 s}{(2\pi)^4} e^{il\cdot s} \ 
({\cal D}^{(n)}_x G_{v} \pm G_{v} {\cal D}^{(n)\dag}_{y}),
\eeq
for $n=0,1,2$.
The expansion of the gauge field $A_{\mu}$ in $\d_X$ reads
\beq
\label{eq:A_ex}
A_{\mu}(x) \approx A_{\mu}(X) + \frac{1}{2}(s \cdot \partial^X) A_{\mu}(X)
+ \frac{1}{8}(s \cdot \partial^X)^2 A_{\mu}(X).
\eeq
Combined with the contributions from the Wilson loop in Eq.~(\ref{eq:covariant}), 
all the terms involving the gauge field are expressed by 
the gauge-invariant field strength $F_{\mu \nu}$ at the end.
Then the third term in the right-hand side of Eq.~(\ref{eq:A_ex}) 
will be irrelevant eventually when $\d^X_{\rho} F_{\mu \nu} = 0$.
Renaming the kinetic residual momentum 
$\tilde l^{\mu} = l^{\mu} - A^{\mu}$ as $l^{\mu}$,
we have
\begin{subequations}
\begin{gather}
I_+^{(0)} = 2(l_0 - l_{\parallel}) G_{v}, \qquad
I_-^{(0)} = iv^{\mu} (g_{\mu 0} \d_t - F_{\mu \nu} \partial^{\nu}_l) G_{v}, 
\\
I_+^{(1)} = \frac{1}{\mu}(-l_{\perp}^2 + {\bf B} \cdot {\bf v}) G_{v},
\qquad
I_-^{(1)} = \frac{i}{\mu} l_{\perp}^{\mu} (g_{\mu 0} \d_t - F_{\mu \nu} \partial^{\nu}_l) G_{v},
\\
\label{eq:I^2}
I_+^{(2)} = \frac{1}{\mu^2}
[l_{\parallel}(l_{\perp}^2  - {\bf B} \cdot {\bf v}) + {\bf B} \cdot {\bf l}_{\perp} 
+ ({\bf E} \times {\bf l}) \cdot {\bf v}] G_{v},
\nonumber \\
I_-^{(2)} = -\frac{i}{2\mu^2}
\left[ 2 l_{\parallel} l_{\perp}^{\mu} - \frac{1}{2} (l_{\perp}^2 
- {\bf B} \cdot {\bf v}) \bar v^{\mu}
- \epsilon^{ijk}v^k \bar v_{\sigma} F^{i \sigma} g^{\mu j} \right] 
(g_{\mu 0} \d_t - F_{\mu \nu} \partial^{\nu}_l) G_{v}, 
\end{gather}
\end{subequations}
where $\d_l$ is the derivative with respect to the residual momentum $l$, and
$l^0 = l_{\parallel} + O(l^2/\mu)$ is used in Eq.~(\ref{eq:I^2}).

From the equation ${\cal D}_x G_{v}(x,y) + G_{v}(x,y){\cal D}_y^{\dag}=0$, 
we obtain the on-shell condition.
Considering the projection conditions (\ref{eq:projection}), $G_{v}$ can be written as
\beq
\label{eq:distribution}
G_{v} = 2\pi {P}_+({\bf v})
\delta \left(l_0- l_{\parallel} - \frac{l_{\perp}^2- {\bf B} \cdot {\bf v}}{2\mu}
+ \frac{l_{\parallel}(l_{\perp}^2  - {\bf B} \cdot {\bf v}) + {\bf B} \cdot {\bf l}_{\perp}}{2\mu^2}
\right) n_{l},
\eeq
where $n_l(X)$ is the distribution function expressed 
by the residual momentum $l$. Recalling
\begin{gather}
p = \mu + l_{\parallel} + \frac{l_{\perp}^2}{2\mu} - 
\frac{l_{\perp}^2 l_{\parallel}}{2\mu^2} + O\left( \frac{1}{\mu^3} \right), \\
\frac{{\bf B} \cdot \hat {\bf p}}{2p} 
= \frac{{\bf B} \cdot {\bf v}}{2\mu} + \frac{{\bf B} \cdot {\bf l}_{\perp} - 
l_{\parallel}{\bf B} \cdot {\bf v}}{2\mu^2} + O\left(\frac{1}{\mu^3} \right),
\end{gather}
the dispersion relation in the delta function of Eq.~(\ref{eq:distribution}) 
is equivalent to the condition $p^0 = \epsilon_{\bf p}$ with
\beq
\label{eq:dispersion_derivative}
\epsilon_{\bf p} = p - \frac{{\bf B} \cdot \hat {\bf p}}{2p},
\eeq
which indeed coincides with Eq.~(\ref{eq:dispersion_Lorentz}).
Accordingly, the distribution function $n_l$ can be replaced
by $n_{\bf p}$ in terms of the original momentum ${\bf p}$.

On the other hand, from the equation ${\cal D}_x G_{v}(x,y) - G_{v}(x,y){\cal D}_y^{\dag} = 0$, 
we obtain the transport equation for $n_{l}(X)$. Using
\beq
\hat {\bf p} = {\bf v} + \frac{{\bf l}_{\perp}}{\mu} 
- \frac{l_{\perp}^2 {\bf v} + 2 l_{\parallel} {\bf l}_{\perp}}{2\mu^2}
+ O \left(\frac{l^3}{\mu^3} \right),
\eeq
the transport equation can also be expressed by the original momentum ${\bf p}$.
We end up with the transport equation
\begin{align}
\label{eq:transport}
& \left(1 + \frac{{\bf B} \cdot \hat {\bf p}}{2\mu^2} \right) \dot n_{\bf p}
+\left[ ({\bf E} + \hat {\bf p} \times {\bf B})
+ ({\bf E} \cdot {\bf B}) \frac{\hat {\bf p}}{2\mu^2} \right] 
\cdot \frac{\d n_{\bf p}}{\d {\bf p}}=0.
\end{align}
Equation (\ref{eq:transport}) indeed agrees with the homogeneous limit 
of the kinetic theory (\ref{eq:kinetic}) to $O(\epsilon_1^2 \epsilon_3^2)$ 
if we identify $\bm{\Omega}_{\bf p} = {\hat {\bf p}}/(2p^2)$ at $p=\mu$.
Therefore, we have found that the Berry curvature corrections in the kinetic theory 
emerge as the higher-order corrections in $1/\mu$ to the usual Vlasov equation.
Once the kinetic equation (\ref{eq:transport}) is obtained, 
the relation of triangle anomalies (\ref{eq:anomaly}) follows, as
we have seen in Sec.~\ref{sec:kinetic}.

\subsection{Particle number density and current}
Here we consider the particle number density and current 
for right-handed fermions without reference to 
the kinetic theory derived above.
To see the chiral magnetic effect, we need to
consider the number current to $O(\epsilon_1 \epsilon_3)$ 
in the high density effective theory.
Our discussion in this subsection is applicable to  
inhomogeneous electromagnetic fields.

By definition, the number density of right-handed fermions 
consists of four parts:
\begin{align}
n 
&= \langle \psi_{+v}^{\dag} \psi_{+v} \rangle
+ \langle \psi_{+v}^{\dag} \psi_{-v} \rangle
+ \langle \psi_{-v}^{\dag} \psi_{+v} \rangle
+ \langle \psi_{-v}^{\dag} \psi_{-v} \rangle 
\nonumber \\  
& \equiv n_{++} + n_{+-} + n_{-+} + n_{--},
\end{align}
where $n_{++}$ is given by
\beq
n_{++} = \int \frac{d^4p}{(2\pi)^4} \tr G_{v} 
= \int \frac{d^3 p}{(2\pi)^3} n_{l},
\eeq
and $n_{+-}=n_{-+}=0$ because of the property of projectors, 
$P_+ ({\bf v}) P_- ({\bf v}) = 0$.
In order to express $n_{--}$ in terms of $\psi_{+v}$, 
we use Eq.~(\ref{eq:relation+-}) which relates $\psi_{-v}$ to $\psi_{+v}$. 
Then $n_{--}$ is given by
\beq
n_{--} = \frac{1}{4\mu^2}
\int \frac{d^4p}{(2\pi)^4} \tr ({\not \! \! D}_{\perp} G_{v} {\not \! \! D}_{\perp}^{\dag})
= \int \frac{d^3 p}{(2\pi)^3} \frac{{\bf B} \cdot {\bf v}}{2\mu^2} n_{l}.
\eeq
Using the momentum ${\bf p}$, we have in total 
\beq
\label{eq:n_micro}
n = \int \frac{d^3 p}{(2\pi)^3} 
\left(1 + \frac{{\bf B} \cdot \hat {\bf p}}{2\mu^2} \right) n_{\bf p}.
\eeq
This is the number density including the Berry curvature correction, 
Eq.~(\ref{eq:n}).

Similarly, the number current of right-handed fermions is decomposed as
\begin{align}
{\bf j}_R 
& = \langle \psi_{+v}^{\dag} {\bm \sigma} \psi_{+v} \rangle
+ \langle \psi_{+v}^{\dag} {\bm \sigma} \psi_{-v} \rangle
+ \langle \psi_{-v}^{\dag} {\bm \sigma} \psi_{+v} \rangle
+ \langle \psi_{-v}^{\dag} {\bm \sigma} \psi_{-v} \rangle \nonumber \\
& \equiv {\bf j}_{++} + {\bf j}_{+-} + {\bf j}_{-+} + {\bf j}_{--},
\end{align}
where ${\bf j}_{++}$ is given by
\beq
{\bf j}_{++} &=&  
\int \frac{d^4p}{(2\pi)^4} \tr({\bm \sigma} G_{v}) 
= \int \frac{d^3 p}{(2\pi)^3}
{\bf v} n_{l}.
\eeq
Using Eq.~(\ref{eq:relation+-}), summation of ${\bf j}_{+-}$ and ${\bf j}_{-+}$
can be written as
\beq
\label{eq:j+-}
{j}^i_{+-} + {j}^i_{-+} = \frac{i}{2\mu}
\int \frac{d^4p}{(2\pi)^4} 
\tr[(D^i_{x \perp}+i \epsilon^{ijk} v^k D_x^j) G_{v}
- G_{v}(D^{\dag i}_{y \perp} - i \epsilon^{ijk} v^k D_y^{\dag j})],
\eeq
while ${\bf j}_{--} \sim 1/\mu^2$ is higher order in $1/\mu$ and 
is negligible to the order under consideration.
One can then rewrite Eq.~(\ref{eq:j+-}) by changing the coordinates from $(x,y)$
to the center-of-mass and relative coordinates $(X,s)$ and 
performing the Wigner transformation in a gauge-covariant way.
One finds
\beq
{j}^i_{+-} + {j}^i_{-+} 
&=& \int \frac{d^3p}{(2\pi)^3} \frac{1}{2\mu}
\left[-\epsilon^{ijk}{v^j}\frac{\d n_{l}}{\d X^k}
+  ({\bf B} \cdot {\bf v}) \frac{\d n_{l}}{\d l^i}
- B^i \left({\bf v} \cdot \frac{\d n_{l}}{\d {\bf l}} \right)
\right].
\eeq
Putting them together and writing in terms of the momentum ${\bf p}$, 
we arrive at
\beq
\label{eq:j_micro}
{j}_R^i &=& 
\int \frac{d^3p}{(2\pi)^3} \left[\frac{\d \epsilon_{\bf p}}{\d p^i} n_{\bf p}
- B^i \left(\frac{\hat {\bf p}}{2\mu} \cdot \frac{\d n_{\bf p}}{\d {\bf p}}\right)
- \epsilon^{ijk} \frac{\hat p^j}{2\mu} \frac{\d n_{\bf p}}{\d X^k} \right], 
\eeq
where we used the dispersion relation (\ref{eq:dispersion}).
This is the same form as the current (\ref{eq:j_sp}), 
including the chiral magnetic effect and the inhomogeneous term; 
the Berry curvature corrections in Eq.~(\ref{eq:j_sp}) microscopically 
originate from the mixing between $\psi_+$ and $\psi_-$.

\section{Correlation functions}
\label{sec:correlation}
In this section we compute the one-loop polarization tensor 
in the presence of chiral fermions at finite chemical potential $\mu$
at zero temperature using the perturbation theory and the kinetic theory 
constructed in Sec.~\ref{sec:kinetic}. 
We confirm that both calculations give the same result, 
not only to the leading order but also to the next-to-leading order in $1/\mu$. 
In particular, we find the kinetic theory with Berry curvature corrections 
reproduces the parity-odd polarization tensor beyond the leading-order
hard dense loop approximation in the perturbation theory.

\subsection{Perturbation theory}
\label{sec:corr_pert}
We first compute the parity-even and parity-odd one-loop polarization 
tensors in the perturbation theory under the hard dense loop approximation.%
\footnote{The parity-odd hard dense loop action was previously derived 
in Ref.~\cite{Laine:2005bt}.}
The time-ordered Dirac fermion propagator at finite chemical potential $\mu$ 
and zero temperature is given by
\begin{align}
S(x,y)=\langle T \psi(x) \bar \psi(y) \rangle =
\int \frac{d^3p}{(2\pi)^3} \frac{\gamma \cdot p}{2p_0}
\left[\theta(x^0-y^0)(\alpha_p e^{-ip(x-y)} + \bar \beta_p e^{ip(x-y)}) \right.
\nonumber \\
\left. \qquad - \theta(y^0-x^0)(\beta_p e^{-ip(x-y)} + \bar \alpha_p e^{ip(x-y)}) \right],
\end{align}
where $\alpha_p = \theta(p_0 - \mu)$, 
$\beta_p = \theta(\mu - p_0)$,
$\bar \alpha_p = 1$, and $\bar \beta_p = 0$. 

The one-loop polarization tensor in the presence of chiral fermions is then
\beq
\Pi^{\mu \nu}(x-y)=\frac{1}{2}\tr[(1 + \gamma_5)\gamma^{\mu}S(x,y)\gamma^{\nu}S(y,x)].
\eeq
In the momentum space, it is given by 
(see Ref.~\cite{Efraty:1992pd} for the case of Dirac fermions)
\beq
\Pi^{\mu \nu}(k) 
&=& \frac{1}{2} \int \frac{d^3 q}{(2\pi)^3}
\frac{1}{2p_0}\frac{1}{2q_0}
\left[ T^{\mu \nu}(p,q) 
\left(\frac{\alpha_p \beta_q}{p_0-q_0-k_0-i\epsilon} 
- \frac{\alpha_q \beta_p}{p_0-q_0-k_0+i\eta} \right) \right.
\nonumber \\
& &
\qquad \qquad \qquad \qquad + T^{\mu \nu}(p,\bar q) 
\left(\frac{\alpha_p \bar \alpha_q}{p_0+q_0-k_0-i\epsilon} 
- \frac{\beta_p \bar \beta_q}{p_0+q_0-k_0+i\eta} \right) 
\nonumber \\
& &
\qquad \qquad \qquad \qquad + T^{\mu \nu}(\bar p, q) 
\left(\frac{\bar \alpha_p \alpha_q}{p_0+q_0+k_0-i\eta} 
- \frac{\bar \beta_p \beta_q}{p_0+q_0+k_0+i\epsilon} \right) 
\nonumber \\
& & \left.
\qquad \qquad \qquad \qquad + T^{\mu \nu}(\bar p, \bar q) 
\left(\frac{\bar \alpha_p \bar \beta_q}{p_0-q_0+k_0-i\eta} 
- \frac{\bar \beta_p \bar \alpha_q}{p_0-q_0+k_0+i\epsilon} \right) \right], 
\eeq
where
$p=(p_0, {\bf p})$, $\bar p=(p_0, -{\bf p})$, $q=(q_0, {\bf q})$,
$\bar q=(q_0, -{\bf p})$, $p_0 =|{\bf p}|$, $q_0 =|{\bf q}|$,
${\bf p}={\bf q} + {\bf k}$, and
$T^{\mu \nu}(p,q)=\tr[(1 + \gamma_5)\gamma^{\mu}{\not \! p} \gamma^{\nu} {\not \! q}]$.
An infinitesimal quantity $\eta$ takes $\eta=\epsilon$ for the time-ordered function 
$\Pi^{\mu \nu}_T$, and $\eta=-\epsilon$ for the retarded function $\Pi^{\mu \nu}_R$.

Substituting explicit expressions of the distribution functions,
we find the $\mu$-dependent part of the retarded function 
(hereafter we suppress the index ``R"), 
\beq
\Pi^{\mu \nu}_{\pm}(k) &=& \frac{1}{2} \int \frac{d^3 q}{(2\pi)^3}
\frac{1}{2p_0}\frac{1}{2q_0}
\left[[\theta(\mu - q_0) - \theta(\mu - p_0)]\frac{T^{\mu \nu}_{\pm}(p,q)}{p_0-q_0-k_0-i\epsilon} \right.
\nonumber \\
& &\left. - \theta(\mu - p_0)\frac{T^{\mu \nu}_{\pm}(p,\bar q)}{p_0+q_0-k_0-i\epsilon}
- \theta(\mu - q_0)\frac{T^{\mu \nu}_{\pm}(\bar p,q)}{p_0+q_0+k_0+i\epsilon}
\right],
\eeq
where $\Pi_{\pm}$ and $T^{\mu \nu}_{\pm}$ denote parity-even and -odd parts,
$T^{\mu \nu}_+(p,q) = \tr(\gamma^{\mu}{\not \! p}\gamma^{\nu}{\not \! q})$
and 
$T^{\mu \nu}_-(p,q) = \tr(\gamma_5 \gamma^{\mu}{\not \! p}\gamma^{\nu}{\not \! q})$.
The parity-even part $\Pi^{\mu \nu}_{+}$ is the leading contribution 
in the hard dense loop approximation, while the parity-odd part 
$\Pi^{\mu \nu}_{-}$ is suppressed compared with $\Pi^{\mu \nu}_{+}$ 
by a factor of $|{\bf k}|/\mu$. 

For completeness, let us first recall the computation of $\Pi^{\mu \nu}_{+}$.
Under the hard dense loop approximation (where $k_0, |{\bf k}|\ll \mu$),
we end up with \cite{Manuel:1995td}
\beq
\Pi^{\mu \nu}_+(k) &=& - \frac{1}{2}\int \frac{d^3 q}{(2\pi)^3} \delta(\mu-|{\bf q}|)
\left(\bar v^{\mu} v^{\nu} + v^{\mu} v^{\nu}
-2\omega \frac{v^{\mu} v^{\nu}}{v \cdot k + i\epsilon} \right)
\nonumber \\
&=& -\frac{\mu^2}{2\pi^2}\left[\delta^{\mu 0} \delta^{\nu 0} -
\omega \int \frac{d{\bf v}}{4\pi} \frac{v^{\mu} v^{\nu}}{v \cdot k + i \epsilon}
\right],
\label{eq:HDL}
\eeq
where $v=(1, {\bf v})$, $\bar v=(1, -{\bf v})$ with ${\bf v}={\bf q}/|{\bf q}|$.

Let us now turn to the parity-odd part $\Pi^{\mu \nu}_-(k)$.
Using $\tr(\gamma_5 \gamma^{\mu} \gamma^{\alpha}
\gamma^{\nu} \gamma^{\beta}) = -4i \epsilon^{\mu \alpha \nu \beta}$,
we have
\beq
\Pi^{\mu \nu}_-(k) = -\frac{1}{2} \int \frac{d^3 q}{(2\pi)^3} \frac{1}{4|{\bf q}|^2}
\left[ {\bf k} \cdot {\bf v} \delta(\mu - |{\bf q}|)
\frac{4i \epsilon^{\mu \nu \alpha \beta}k_{\alpha}v_{\beta}|{\bf q}|}
{k \cdot v + i \epsilon}
- \theta(\mu-|{\bf q}|) {2i\epsilon^{\mu \nu \alpha \beta}
(k_{\alpha} \bar v_{\beta} + \bar k_{\alpha} v_{\beta})}
\right].
\nonumber \\
\eeq
The second term is vanishing while the first term 
remains nonzero only when $(\alpha, \beta)=(0, k)$ and $(k,0)$.
Collecting both contributions, we obtain the expression
\beq
\label{eq:corr_pert1}
\Pi^{i j}_-(k) = \frac{\mu}{4\pi^2} \left(i \epsilon^{ijk}k^k 
+ i \epsilon^{ijk} \omega 
\int \frac{d {\bf v}}{4\pi} \frac{\omega v^k - k^k}{k \cdot v + i\epsilon}
\right),
\eeq
where $i,j,k$ denote the spatial indices
[$\Pi^{\mu \nu}_-(k)$ is vanishing otherwise]. 
Performing the angular integration,
we finally arrive at
\beq
\label{eq:corr_pert2}
\Pi^{i j}_-(k) = \frac{\mu}{4\pi^2} i \epsilon^{ijk}k^k 
\left(1 - \frac{\omega^2}{|{\bf k}|^2} \right)[1 - \omega L(k)],
\eeq
where
\beq
\label{eq:Lindhard}
L(k)=\frac{1}{2|{\bf k}|}\ln \frac{\omega + |\bf k|}{\omega-|\bf k|}.
\eeq

Equation~(\ref{eq:corr_pert2}) reduces to a simple form in 
the static or long wavelength limit:
\begin{eqnarray}
\Pi^{i j}_-(k) = \left\{ \begin{array}{ll}
   \displaystyle{\frac{\mu}{4\pi^2} i \epsilon^{ijk}k^k} & (\omega \ll |{\bf k}|)  \\
   \displaystyle{\frac{\mu}{12\pi^2} i \epsilon^{ijk}k^k} & (\omega \gg |{\bf k}|) \\
\end{array}   \right. .
\end{eqnarray} 
That the latter $\lim_{|{\bf k}|/\omega \rightarrow 0} \Pi^{ij}_-(k)$ is smaller than 
the former $\lim_{\omega/|{\bf k}| \rightarrow 0} \Pi^{ij}_-(k)$ by a factor of 3 is 
consistent with the result of the ``chiral magnetic conductivity" 
in Ref.~\cite{Kharzeev:2009pj}.

\subsection{Kinetic theory with Berry curvature}
\label{sec:corr_kin}
We now compute the same retarded correlation function 
from the kinetic theory (\ref{eq:kinetic}) constructed 
in Sec.~\ref{sec:kinetic} through the linear response theory
\beq
j^{\mu}(x) = \int d^4y \ \Pi^{\mu \nu}(x-y) A_{\nu}(y),
\eeq
or in the momentum space
\beq
\label{eq:LRT}
j^{\mu}(k) = \Pi^{\mu \nu}(k) A_{\nu}(k).
\eeq
Here we are interested in the current induced by a linear-order 
deviation of the gauge field $A_{\mu}$. 
For definiteness, we set up the following power counting scheme:
$A_{\mu} = O(\epsilon)$ and $\d_x = O(\delta)$, where $\epsilon$ and
$\delta$ are small and independent expansion parameters.
Under this counting scheme, we compute the deviation of the distribution 
function $\delta n_{\bf p}$ and the current $j^{\mu}$ to $O(\epsilon \delta)$.  

Remembering that $\d n_{\bf p}/\d {\bf x}$ 
or $\d \epsilon_{\bf p}/\d {\bf x}$ can be
nonvanishing at least when $\d n_{\bf p}/\d {\bf x} = O(\epsilon)$ or 
$\d \epsilon_{\bf p}/\d {\bf x} = O(\epsilon)$ in Eq.~(\ref{eq:kinetic}),
it is sufficient to consider the following kinetic equation of order 
$O(\epsilon \delta^2)$ [or $n_{\bf p}$ of order $O(\epsilon \delta)$],
\beq
\label{eq:simple_kin}
\left(\frac{\d}{\d t} + {\bf v} \cdot \frac{\d}{\d {\bf x}} \right) n_{\bf p}
+\left({\bf E} + {\bf v} \times {\bf B} - 
\frac{\d \epsilon_{\bf p}}{\d {\bf x}} \right) \cdot \frac{\d n_{\bf p}}{\d {\bf p}}=0,
\eeq
in which the ${\bf v} \times {\bf B}$ term does not contribute
since $({\bf v} \times {\bf B}) \cdot {\bf v}=0$.
The distribution function $n_{\bf p}$ is decomposed as
\beq
n_{\bf p} = n_{\bf p}^{(0)} + n_{\bf p}^{(\epsilon)} 
+ n_{\bf p}^{(\epsilon \delta)} + \cdots,
\eeq
where
\beq
\label{eq:initial}
n_{\bf p}^{(0)} =
\theta (\mu- \epsilon_{\bf p}) \simeq \theta(\mu-|{\bf p}|) 
+ \frac{{\bf B} \cdot {\bf v}}{2\mu}\delta(\mu-|{\bf p}|),
\eeq
which follows from the dispersion relation (\ref{eq:dispersion_Lorentz}).
Note that the second term in Eq.~(\ref{eq:initial}) is also $O(\epsilon \delta)$,
but this is separated from $n_{\bf p}^{(\epsilon \delta)}$ in our definition.
In the calculations below, we have to add both contributions at the same order 
of $O(\epsilon \delta)$ [see Eq.~(\ref{eq:order2}) below].

The kinetic equation can be written down at each order as 
\begin{gather}
\label{eq:order1}
\left(\frac{\d}{\d t} + {\bf v} \cdot \frac{\d}{\d {\bf x}} \right) n_{\bf p}^{(\epsilon)}
= {\bf E} \cdot {\bf v} \delta(\mu-|{\bf p}|), \\
\label{eq:order2}
\left(\frac{\d}{\d t} + {\bf v} \cdot \frac{\d}{\d {\bf x}} \right) 
\left( n_{\bf p}^{(\epsilon \delta)} + \frac{{\bf B} \cdot {\bf v}}{2\mu}\delta(\mu-|{\bf p}|) \right)
- {\bf v} \cdot \frac{\d}{\d {\bf x}} \left( \frac{{\bf B} \cdot {\bf v}}{2\mu} \right)
\delta(\mu-|{\bf p}|) = 0.
\end{gather}
The second equation is further simplified to
\beq
\left(\frac{\d}{\d t} + {\bf v} \cdot \frac{\d}{\d {\bf x}} \right) 
n_{\bf p}^{(\epsilon \delta)} = - \frac{\d}{\d t} \left( \frac{{\bf B} \cdot {\bf v}}{2\mu} \right)
\delta(\mu-|{\bf p}|).
\eeq

Using the method of characteristics, we can solve these equations; 
the operator in the left-hand sides of equations $v \cdot \partial_x$
is the time derivative along the characteristic ${\bf v}=d{\bf x}/d t$.
The solutions are given by
\begin{align}
\label{eq:solution1}
n_{\bf p}^{(\epsilon)} 
&= \delta(\mu-|{\bf p}|)\int_0^{\infty} d \tau \ e^{- \eta \tau}
{\bf v}\cdot {\bf E}(x - v \tau),
\\
\label{eq:solution2}
n_{\bf p}^{(\epsilon \delta)} &= 
-\delta(\mu-|{\bf p}|)\int_0^{\infty} d \tau \ e^{- \eta \tau}
\frac{1}{{2\mu}} {\bf v} \cdot \dot {\bf B}(x -  v \tau),
\end{align}
where $\eta$ is a small positive parameter which ensures
${\bf E}(t \rightarrow -\infty, {\bf x})\rightarrow 0$ and
${\bf B}(t \rightarrow -\infty, {\bf x})\rightarrow 0$.

Now let us compute the current defined in Eq.~(\ref{eq:j}).
The current can be written down at the orders of $O(\epsilon)$ and 
$O(\epsilon \delta)$, respectively, as
\beq
\label{eq:j1}
j^{\mu(\epsilon)}(x) &=& 
\int \frac{d^3 p}{(2\pi)^3} 
v^{\mu} n_{\bf p}^{(\epsilon)},
\\
\label{eq:j2}
j^{i(\epsilon \delta)}(x) &=& 
\int \frac{d^3 p}{(2\pi)^3}
\left[v^i n_{\bf p}^{(\epsilon \delta)}  + \frac{B^i}{2\mu} \delta(\mu-|{\bf p}|) 
- \epsilon^{ijk} \frac{v^j}{2\mu}
\frac{\d n_{\bf p}^{(\epsilon)}}{\d x^k} \right],
\eeq
where $v^{\mu}=(1,{\bf v})$. The zeroth component of the four current
of order $O(\epsilon \delta)$, i.e., the number density $n^{(\epsilon \delta)}(x)$, 
is found to vanish after the angular integration.

First consider the four current $j^{\mu(\epsilon)}$.
Substituting Eq.~(\ref{eq:solution1}) into Eq.~(\ref{eq:j1}), 
the current reads
\beq
j^{\mu(\epsilon)}(x) = \int \frac{d^3 p}{(2\pi)^3} v^{\mu} \delta(\mu-|{\bf p}|)
\int_0^{\infty} d \tau \ e^{- \eta \tau}
{\bf v} \cdot {\bf E}(x - v \tau).
\eeq
Using the useful formula
\beq
\label{eq:useful_formula}
\int d^4 x \ e^{ik \cdot x} \int_0^{\infty} d \tau \
e^{-\eta \tau} f(x-v \tau)
=\frac{if(k)}{v \cdot k + i \eta},
\eeq
and the linear response theory (\ref{eq:LRT}), we obtain the retarded 
parity-even polarization tensor \cite{Silin}
\beq
\Pi^{\mu \nu}_+(k)= -\frac{\mu^2}{2\pi^2}\left[\delta^{\mu 0} \delta^{\nu 0} -
\omega \int \frac{d{\bf v}}{4\pi} \frac{v^{\mu} v^{\nu}}{v \cdot k + i \epsilon}
\right],
\eeq
which agrees with Eq.~(\ref{eq:HDL}) derived from the
perturbation theory.

We then turn to the subleading three current $j^{i(\epsilon \delta)}$.
Substituting Eqs.~(\ref{eq:solution1}) and (\ref{eq:solution2})
into (\ref{eq:j2}), and using the formula (\ref{eq:useful_formula}),
the current reads
\beq
j^{i(\epsilon \delta)}(k) =
\int \frac{d^3 p}{(2\pi)^3} \delta(\mu-|{\bf p}|) 
\left[ -\frac{i \epsilon^{klm} v^i v^k \omega k^l A^m}
{2\mu (v \cdot k + i\epsilon)}
+ \frac{i \epsilon^{ijk}k^j A^k}{2\mu} 
+ \frac{i \epsilon^{ikl} v^k k^l
\left[\omega({\bf v \cdot A}) - ({\bf v \cdot k}) A_0\right]}
{2\mu (v \cdot k + i\epsilon)}
\right],
\nonumber \\ 
\eeq
among which the $A_0$ term in the brackets vanishes after the 
angular integration.
Using the linear response theory (\ref{eq:LRT}), we obtain the parity-odd
polarization tensor
\beq
\label{eq:corr_kin1}
\Pi^{i j}_-(k) = \frac{\mu}{4\pi^2} \left[i \epsilon^{ijk}k^k 
+ i \omega 
\int \frac{d {\bf v}}{4\pi} \frac{
(\epsilon^{jkl}v^i -
\epsilon^{ikl}v^j) v^k k^l}{v \cdot k + i\epsilon}
\right].
\eeq
After the angular integration, this reduces to the form:
\beq
\label{eq:corr_kin2}
\Pi^{i j}_-(k) = \frac{\mu}{4\pi^2} i \epsilon^{ijk}k^k 
\left(1 - \frac{\omega^2}{|{\bf k}|^2} \right)[1 - \omega L(k)],
\eeq
where $L(k)$ is defined in Eq.~(\ref{eq:Lindhard}).
This is equivalent to Eq.~(\ref{eq:corr_pert2}) derived 
from the perturbation theory, which confirms that the physics 
in the next-to-leading order hard dense loop approximation can 
be described by the kinetic theory with Berry curvature corrections.
Note that contributions of the inhomogeneous term in the current (\ref{eq:j_sp}) 
and the magnetic moment in Eq.~(\ref{eq:dispersion_Lorentz}) are necessary for
the matching of the correlation functions.

\section{Conclusion}
\label{sec:conclusion}
In this paper, we have shown a way to bridge between quantum field theories 
and the kinetic theory with Berry curvature corrections that exhibits
triangle anomalies and the chiral magnetic effect.
The field theoretic procedure to derive such a kinetic theory developed in this paper 
can, in principle, be generalized to higher order in gauge fields and/or derivatives.
We have also computed the parity-odd correlation function
using this kinetic theory, which was found to agree with the perturbative result 
beyond the leading-order hard dense loop approximation.

It should be remarked that our derivation of the kinetic theory from 
underlying quantum field theories is limited to low temperature region $T \ll \mu$ 
where the Fermi surface is well defined and the high density effective theory is applicable;
a generalization to higher temperature regime would be desirable.
We also remark that our formulation based on the Berry curvature 
is not manifestly Lorentz covariant by construction.
It might be possible to formulate the kinetic theory in a Lorentz 
covariant way similarly to the usual Vlasov equation.
Without referring to the Berry curvatures,
such a solution of the kinetic equation in the hydrodynamic regime 
was obtained in Ref.~\cite{Gao:2012ix} which also reproduces the chiral 
vortical effect \cite{Vilenkin:1979ui, Kharzeev:2007tn, Son:2009tf, Landsteiner:2011cp, Landsteiner:2011iq}.

The inclusion of collision terms in the kinetic theory (\ref{eq:kinetic})
is straightforward, which gives the modified Boltzmann equation 
taking into account the anomalous effects.
We hope that our work motivates numerical applications of
the new kinetic equation (\ref{eq:kinetic}) with or without collisions 
in various systems such as hot and dense quark matter, neutrino gas, 
the early Universe at large lepton chemical potential, and doped Weyl semimetals.

\acknowledgments
N.Y. is supported by JSPS Research Fellowships for Young Scientists. 
This work is supported in part by DOE Grant No.\ DE-FG02-00ER41132.

\emph{Note added.}---While this work was being completed, we were informed 
of Ref.~\cite{Chen:2012ca} where the authors have also derived the kinetic theory 
with Berry curvature corrections from quantum field theories using a Wigner function.


\begin{thebibliography}{99}

\bibitem{Landau_kinetics}
 L.~D.~Landau and E.~M.~Lifshitz, {\it Physical Kinetics}
(Pergamon, New York, 1981).

\bibitem{Adler}
  S.~Adler,
  Phys.\ Rev.\ {\bf 177}, 2426 (1969).

\bibitem{BellJackiw}
  J.~S.~Bell and R.~Jackiw,
  Nuovo Cimento A {\bf 60}, 47 (1969).

\bibitem{Son:2012wh} 
  D.~T.~Son and N.~Yamamoto,
  Phys.\ Rev.\ Lett.\  {\bf 109}, 181602 (2012)
  [arXiv:1203.2697 [cond-mat.mes-hall]].

\bibitem{Kirilin:2012sd} 
  V.~P.~Kirilin, Z.~V.~Khaidukov, and A.~V.~Sadofyev,
  Phys.\ Lett.\ B {\bf 717}, 447 (2012)
  [arXiv:1203.6612 [cond-mat.mes-hall]].

\bibitem{Zahed:2012yu} 
  I.~Zahed,
  Phys.\ Rev.\ Lett.\  {\bf 109}, 091603 (2012)
  [arXiv:1204.1955 [hep-th]].

\bibitem{Son:2012bg}
  D.~T.~Son and B.~Z.~Spivak,
  arXiv:1206.1627 [cond-mat.mes-hall].  

\bibitem{Gorsky:2012gi} 
  A.~Gorsky and A.~V.~Zayakin,
  JHEP {\bf 1302}, 124 (2013)
  [arXiv:1206.4725 [hep-th]].

\bibitem{Stephanov:2012ki} 
  M.~A.~Stephanov and Y.~Yin,
  Phys.\ Rev.\ Lett.\  {\bf 109}, 162001 (2012)
  [arXiv:1207.0747 [hep-th]].

\bibitem{Loganayagam:2012pz} 
  R.~Loganayagam and P.~Surowka,
  JHEP {\bf 1204}, 097 (2012)  
  [arXiv:1201.2812 [hep-th]].  
  
\bibitem{Gao:2012ix} 
  J.~-H.~Gao, Z.~-T.~Liang, S.~Pu, Q.~Wang and X.~-N.~Wang,
  Phys.\ Rev.\ Lett.\  {\bf 109}, 232301 (2012)
  [arXiv:1203.0725 [hep-ph]].

\bibitem{Berry}
  M.~V.~Berry, Proc.~R.~Soc.~Lond. A {\bf 392}, 45 (1984).

\bibitem{Xiao:2010}
  D.~Xiao, M.-C.~Chang, and Q.~Niu, 
  Rev.\ Mod.\ Phys.\ {\bf 82}, 1959 (2010) 
  [arXiv:0907.2021 [cond-mat.mes-hall]]. 

\bibitem{Vilenkin:1980fu} 
  A.~Vilenkin,
  Phys.\ Rev.\ D {\bf 22}, 3080 (1980).  

\bibitem{Nielsen:1983rb} 
  H.~B.~Nielsen and M.~Ninomiya,
  Phys.\ Lett.\ B {\bf 130}, 389 (1983).  

\bibitem{Alekseev:1998ds} 
  A.~Y.~Alekseev, V.~V.~Cheianov, and J.~Frohlich,
  Phys.\ Rev.\ Lett.\  {\bf 81}, 3503 (1998)  [cond-mat/9803346].  

\bibitem{Fukushima:2008xe}
  K.~Fukushima, D.~E.~Kharzeev, and H.~J.~Warringa,
  Phys.\ Rev.\  D {\bf 78}, 074033 (2008).
  [arXiv:0808.3382 [hep-ph]].
  
\bibitem{Erdmenger:2008rm} 
  J.~Erdmenger, M.~Haack, M.~Kaminski, and A.~Yarom,
  JHEP {\bf 0901}, 055 (2009)  [arXiv:0809.2488 [hep-th]].  

\bibitem{Banerjee:2008th} 
  N.~Banerjee, J.~Bhattacharya, S.~Bhattacharyya, S.~Dutta, R.~Loganayagam, and P.~Surowka,
  JHEP {\bf 1101}, 094 (2011)  [arXiv:0809.2596 [hep-th]].  
  
\bibitem{Son:2009tf}
  D.~T.~Son and P.~Surowka,
  Phys.\ Rev.\ Lett.\  {\bf 103}, 191601 (2009)  [arXiv:0906.5044 [hep-th]].  

\bibitem{Banerjee:2012iz} 
  N.~Banerjee, J.~Bhattacharya, S.~Bhattacharyya, S.~Jain, S.~Minwalla, and T.~Sharma,
  JHEP {\bf 1209}, 046 (2012)
  [arXiv:1203.3544 [hep-th]].
  
\bibitem{Jensen:2012jy} 
  K.~Jensen,
  Phys.\ Rev.\ D {\bf 85}, 125017 (2012)
  [arXiv:1203.3599 [hep-th]].

\bibitem{Kharzeev:2010gr} 
  D.~E.~Kharzeev and D.~T.~Son,
  Phys.\ Rev.\ Lett.\  {\bf 106}, 062301 (2011)  [arXiv:1010.0038 [hep-ph]].  

\bibitem{Vishwanath}
  X.~Wan, A.~M.~Turner, A.~Vishwanath, and S.~Y.~Sav\-ra\-sov,
  Phys.\ Rev.\ {\bf B} 83, 205101 (2011) [arXiv:1007.0016 [cond-mat.str-el]].

\bibitem{BurkovBalents}
  A.~A.~Burkov and L.~Balents,
  Phys.\ Rev.\ Lett.\ {\bf 107}, 127205 (2011) [arXiv:1105.5138 [cond-mat.mes-hall]].

\bibitem{Xu-chern}
  G.~Xu, H.~Weng, Z.~Wang, X.~Dai, and Z.~Fang,
  Phys.\ Rev.\ Lett. {\bf 107}, 186806 (2011) [arXiv:1106.3125 [cond-mat.mes-hall]].

\bibitem{Blaizot:2001nr} 
  J.~-P.~Blaizot and E.~Iancu,
  Phys.\ Rept.\  {\bf 359}, 355 (2002)  [hep-ph/0101103].  
  
\bibitem{ShindouBalents}
  R.~Shindou and L.~Balents,
  Phys.\ Rev.\ B {\bf 77}, 035110 (2008)
  [arXiv:0706.4251 [cond-mat.str-el]].

\bibitem{WongTserkovnyak}
  C.~H.~Wong and Y.~Tserkovnyak,
  Phys.\ Rev.\ B {\bf 84}, 115209 (2011) 
  [arXiv:1102.1121 [cond-mat.mes-hall]].

\bibitem{Hong:1998tn}
  D.~K.~Hong,
  Phys.\ Lett.\  B {\bf 473}, 118 (2000)
  [arXiv:hep-ph/9812510];
  Nucl.\ Phys.\  B {\bf 582}, 451 (2000)
  [arXiv:hep-ph/9905523].

\bibitem{Schafer:2003jn}
  T.~Sch\"afer,
  Nucl.\ Phys.\  A {\bf 728}, 251 (2003)
  [arXiv:hep-ph/0307074].

\bibitem{Xiao:2005}
  D.~Xiao, J.~Shi, and Q.~Niu,
  Phys.\ Rev.\ Lett.\ {\bf 95}, 137204 (2005) [cond-mat/0502340].

\bibitem{Duval:2005}
  C.~Duval, Z.~Horv\'ath, P.~A.~Horv\'athy, L.~Martina, and P.~Stichel,
  Mod.\ Phys.\ Lett.\ B {\bf 20}, 373 (2006) [cond-mat/0506051].

\bibitem{Manuel:1995td} 
  C.~Manuel,
  Phys.\ Rev.\ D {\bf 53}, 5866 (1996)  [hep-ph/9512365].  

\bibitem{Landau}
  L.D.~Landau,
  Sov.\ Phys.\ JETP {\bf 3}, 920 (1957);
  ibid. {\bf 8}, 70 (1959).

\bibitem{Baym:1975va}
  G.~Baym and S.~A.~Chin,
  Nucl.\ Phys.\  A {\bf 262}, 527 (1976).

\bibitem{Migdal}
  A.~B.~Migdal, {\it Theory of Finite Fermi Systems and Applications to Finite Nuclei}
  (Interscience, London, 1967).

\bibitem{Bentz:1985qh} 
  W.~Bentz, A.~Arima, H.~Hyuga, K.~Shimizu, and K.~Yazaki,
  Nucl.\ Phys.\ A {\bf 436}, 593 (1985).  

\bibitem{Laine:2005bt} 
  M.~Laine,
  JHEP {\bf 0510}, 056 (2005)
  [hep-ph/0508195].

\bibitem{Efraty:1992pd}
  R.~Efraty and V.~P.~Nair,
  Phys.\ Rev.\ D {\bf 47}, 5601 (1993)  [hep-th/9212068];  
  R.~Jackiw and V.~P.~Nair,
  Phys.\ Rev.\ D {\bf 48}, 4991 (1993)  [hep-ph/9305241].  
  
\bibitem{Kharzeev:2009pj} 
  D.~E.~Kharzeev and H.~J.~Warringa,
  Phys.\ Rev.\ D {\bf 80}, 034028 (2009)  [arXiv:0907.5007 [hep-ph]].  

\bibitem{Silin}
  V.~P.~Silin, Sov.\ Phys.\ JETP {\bf 11}, 1136 (1960).

\bibitem{Vilenkin:1979ui}
  A.~Vilenkin,
  Phys.\ Rev.\ D {\bf 20}, 1807 (1979).  

\bibitem{Kharzeev:2007tn} 
  D.~Kharzeev and A.~Zhitnitsky,
  Nucl.\ Phys.\ A {\bf 797}, 67 (2007)  [arXiv:0706.1026 [hep-ph]].  

\bibitem{Landsteiner:2011cp} 
  K.~Landsteiner, E.~Megias, and F.~Pena-Benitez,
  Phys.\ Rev.\ Lett.\  {\bf 107}, 021601 (2011)  [arXiv:1103.5006 [hep-ph]].  

\bibitem{Landsteiner:2011iq} 
  K.~Landsteiner, E.~Megias, L.~Melgar, and F.~Pena-Benitez,
  JHEP {\bf 1109}, 121 (2011)  [arXiv:1107.0368 [hep-th]].  

\bibitem{Chen:2012ca} 
  J.~-W.~Chen, S.~Pu, Q.~Wang, and X.~-N.~Wang,
  arXiv:1210.8312 [hep-th].

\end{thebibliography}
\end{document}